\pdfoutput=1

\documentclass{PoS}

\usepackage{booktabs}
\usepackage{url}
\usepackage{amssymb,amsmath,amsfonts}
\usepackage[utf8]{inputenc}
\usepackage{upgreek}
\usepackage[implicit=false,hidelinks]{hyperref}
\usepackage{color}
\hypersetup{
    colorlinks,
    linkcolor={blue},
    citecolor={blue},
    urlcolor={blue}
}

\usepackage{cite}

\newcommand{\e}{\mathrm{e}}
\renewcommand{\d}{\mathrm d}

\let\originalleft\left
\let\originalright\right
\renewcommand{\left}{\mathopen{}\mathclose\bgroup\originalleft}
\renewcommand{\right}{\aftergroup\egroup\originalright}

\title{Topological Susceptibility to High Temperatures via Reweighting}

\ShortTitle{Topological Susceptibility to High Temperatures via Reweighting}

\author{\speaker{P. Thomas Jahn}, Guy D. Moore, and Daniel Robaina\\
        Institut f\"ur Kernphysik (Theoriezentrum), Technische Universit\"at Darmstadt, Schlossgartenstr. 2, D-64289 Darmstadt, Germany\\
        E-mail: \email{\{tjahn,guymoore,robaina\}@theorie.ikp.physik.tu-darmstadt.de}}
\abstract{
We measure the topological susceptibility of quenched QCD on the lattice at two high temperatures.
For this, we define topology with the help of gradient flow and mitigate the statistical problem of
topology at high temperatures using a reweighting technique. This allows us to enhance tunneling
events between topological sectors and alleviate topological freezing.
We quote continuum extrapolated results for the susceptibility at
$2.5$ and $4.1~T_\mathrm c$ that agree well with the existing literature. We conclude that the
method is feasible and can be extended to unquenched QCD with no conceptual problems.
}

\FullConference{The 36th Annual International Symposium on Lattice Field Theory - LATTICE2018\\
		22-28 July, 2018\\
		Michigan State University, East Lansing, Michigan, USA.}

\begin{document}

\section{Introduction}

The axion \cite{Weinberg:1977ma,Wilczek:1977pj} is a hypothetical light scalar particle that could explain the origin of dark matter
and solve the strong CP problem at the same time. It is introduced via the Peccei-Quinn
mechanism \cite{Peccei:1977hh,Peccei:1977ur} that explains why the CP violating $\theta_\mathrm{QCD}$ term
in the QCD Lagrangian is small. The corresponding particle on the other hand could
be a candidate for dark matter in the Universe. In the last years, the high-energy physics community has put a
lot of effort on the QCD axion, both in experiment and theory.

Given the interest on axions, theoretical predictions of its properties would be valuable.
According to the Peccei-Quinn theory, dark-matter axion production is sensitive to the temperature dependence
of the QCD topological susceptibility
\begin{align}
\chi(T) = \int \d^4x \ \left\langle q(x)q(0) \right\rangle = \frac{1}{\beta V} \left\langle Q^2 \right\rangle
\label{eq:chitop}
\end{align}
up to temperatures of about $7~T_\mathrm c$ \cite{Moore:2017ond}, where
\begin{align}
q(x) = \frac{1}{64\pi^2} \epsilon_{\mu\nu\rho\sigma} F_{\mu\nu}^a(x) F_{\rho\sigma}^a(x)
\end{align}
is the topological charge density and $Q=\int \d^4x \ q(x)$ the topological charge.

At low temperatures, the value of the topological susceptibility is well established \cite{diCortona:2015ldu}, while at
high temperatures lattice calculations become very challenging. The main problem is that topologically non-trivial configurations,
the so-called instantons, become very rare as the temperature increases. Therefore, sampling
topological sectors with standard lattice techniques that rely on stochastic approaches fails at high temperatures.
Another problem is topological freezing, which describes a deficiency of update algorithms that
tend to get stuck in topological sectors.

There has been a lot of recent progress in studying topology at high temperatures
\cite{Frison:2016vuc,Berkowitz:2015aua,Taniguchi:2016tjc,%
Bonati:2015vqz,Petreczky:2016vrs,Borsanyi:2015cka,Borsanyi:2016ksw}.
One way \cite{Taniguchi:2016tjc,Borsanyi:2016ksw} to get to high temperatures is to start at a small temperature where
instantons are not rare and differentially work up to high temperatures
by studying fixed topological ensembles. As an alternative, we study topology directly at a given temperature by ameliorating
the sampling problem using a reweighting approach. For simplicity, we constrained ourselves to the
quenched approximation for developing this method, but we see no conceptual problems in applying the
same technique to the unquenched case. Indeed, a very similar approach was recently applied to
$N_\mathrm f = 2+1$ QCD \cite{Bonati:2018blm}. We first presented the method described here in
Ref.~\cite{Jahn:2018dke}, where the reader can find more details.

\section{Reweighting Method}

The difficulty that arises when measuring the topological susceptibility at high temperatures is twofold.
On the one hand, the quantity is physically small, meaning that almost all configurations have trivial topology
and a canonical sample has very little topological information.
On the other hand, typical update algorithms tend to get stuck in topological sectors at fine lattices,
preventing tunneling between topological sectors. The reweighting technique we developed has its roots
in Refs.~\cite{Berg:1991cf,Kajantie:1995kf,Wang:2000fzi}. In the following, we summarize the basic ingredients
of reweighting.

In the path-integral representation, the expectation value of an observable $O$ is given as
\begin{align}
\left\langle O \right\rangle = \frac{\int \mathcal DU \e^{-S_\mathrm W[U]} O[U]}{\int \mathcal DU \e^{-S_\mathrm W[U]}},
\label{eq:contobservable}
\end{align}
where $S_\mathrm W$ is the standard Wilson action and $U$ denotes the gauge links. In lattice gauge theory, this quantity is
approximated by generating lattice configurations distributed according to the probability distribution
\begin{align}
\d P(U) = \frac{\e^{-S_\mathrm W[U]} \mathcal D U}{\int \mathcal DU \e^{-S_\mathrm W[U]}}.
\end{align}
In this way, Eq.~\eqref{eq:contobservable} turns into
\begin{align}
\left\langle O \right\rangle_\mathrm{lat} = \frac 1N \sum_{i=1}^N O_i,
\end{align}
where $i=1,\dots,N$ runs through the sample configurations.

The idea of reweighting is to rewrite Eq.~\eqref{eq:contobservable} as
\begin{align}
\left\langle O \right\rangle = \frac{\int \mathcal DU \e^{-S_\mathrm W[U] + W(\xi)} \e^{-W(\xi)} O[U]}{\int \mathcal DU \e^{-S_\mathrm W[U] + W(\xi)} \e^{-W(\xi)}}
\end{align}
and create a sample of configurations according to the now modified probability distribution
\begin{align}
\d P_\mathrm{rew}(U) = \frac{\e^{-S_\mathrm W[U] +W(\xi)} \mathcal D U}{\int \mathcal DU \e^{-S_\mathrm W[U] + W(\xi)}}
\label{eq:rewprobdist}
\end{align}
that depends on the so-called \emph{reweighting function} $W(\xi)$. In this way, we can artificially modify the weight of configurations
with non-trivial topology. In order to compensate for this modified weight,
we need to redefine the lattice expectation value as
\begin{align}
\left\langle O \right\rangle = \frac{\sum_i^N O_i \e^{-W(\xi_i)}}{\sum_i^N \e^{-W(\xi_i)}}.
\label{eq:rewobservable}
\end{align}
Note that Eqs.~\eqref{eq:contobservable} and \eqref{eq:rewobservable} form a mathematical identity
if the algorithm tends to Eq.~\eqref{eq:rewprobdist}
and $N\to\infty$. The reweighting approach is therefore correct for any choice of the reweighting function.
However, the particular choice is important to improve statistics in the case of topology at high temperatures
as described in the next subsection.
In the following, we use the HMC algorithm to update our configurations and implement the
reweighting part in an additional Metropolis accept/reject step in terms of the change of $W$ after the standard
Metropolis step in terms of the change of the Hamiltonian. For more details of the specific implementation we
refer to Ref.~\cite{Jahn:2018dke}.

\subsection{Reweighting Variable}
The key ingredient in the reweighting approach is the reweighting function $W(\xi)$ depending on the reweighting variable $\xi$.
The idea is to enhance the number of topologically non-trivial configurations by giving them a larger weight, i.e., a large
$W(\xi)$. Consequently, the reweighting variable $\xi$ needs be able to distinguish between different topological sectors.
A natural choice would therefore be the topological charge
\begin{align}
Q =  \frac{1}{64\pi^2} \epsilon_{\mu\nu\rho\sigma} \sum_x \hat F^a_{\mu\nu}(x) \hat F^a_{\rho\sigma}(x),
\end{align}
where for the discretized field strength tensor we use an $\mathcal O\left(a^2\right)$ improved version. However,
the topological charge is badly contaminated by UV fluctuations that add large contributions to the topological
charge. In order to remove those, some amount of gradient flow \cite{Narayanan:2006rf,Luscher:2009eq} is applied
and we define the reweighting variable as
\begin{align}
\xi = Q' \equiv \left| \frac{1}{64\pi^2} \epsilon_{\mu\nu\rho\sigma} \sum_x \left( \hat F^a_{\mu\nu}(x) \hat F^a_{\rho\sigma}(x) \right)_{t'} \right|,
\end{align}
where $t'$ denotes a small amount of gradient flow. With small in this context we mean that UV
fluctuations are removed, while dislocations, which are small concentrations of topological charge
that are the intermediate steps between $Q=0$ and $Q=1$, are still present. Note that at high
temperatures instantons are highly suppressed such that it is sufficient to only regard the $Q=1$
sector. Consequently, $Q'$ distinguishes between $Q=0$ configurations, intermediate
configurations with non-integer $Q$ (dislocations), and $Q=1$ configurations (instantons or calorons).
Note that we here use the absolute value in the definition of $Q'$ making use of the symmetry $Q \to -Q$.

%\subsection{Implementation}
%In this subsection we describe our implementation of the reweighting idea that ensures that configurations are built with
%the probability distribution Eq.~\eqref{eq:rewprobdist}, assuming that the function $W(Q')$ is known. Our update algorithm
%is based on the hybrid Monte Carlo algorithm (HMC). Here we integrate the Hamiltonian equations of motion followed by
%a Metropolis accept/reject step that accepts the new configuration with probability
%\begin{align}
%P_\mathrm{HMC}=\min\left\{ 1,\e^{-\Delta H} \right\},
%\end{align}
%where $\Delta H$ is the change of the Hamiltonian. Reweighting is now incorporated by adding an additional Metropolis
%accept/reject step in terms of the change of the reweighting function: If the HMC accept/reject step accepts the new configuration,
%it is not completely accepted, by we only accept it with probability
%\begin{align}
%P_\mathrm{rew}=\min\left\{ 1,\e^{\Delta W} \right\},
%\end{align}
%where $\Delta W$ indicates the change of the reweighting function.

\subsection{Reweighting Function}
In order to build the reweighting function, we perform a separate Markov chain
where we change $W$ after each trajectory. A second, completely independent
Markov chain is then used with $W$ fixed  to extract physics information.

\begin{figure}
	\centering
	\includegraphics[width=.6\textwidth]{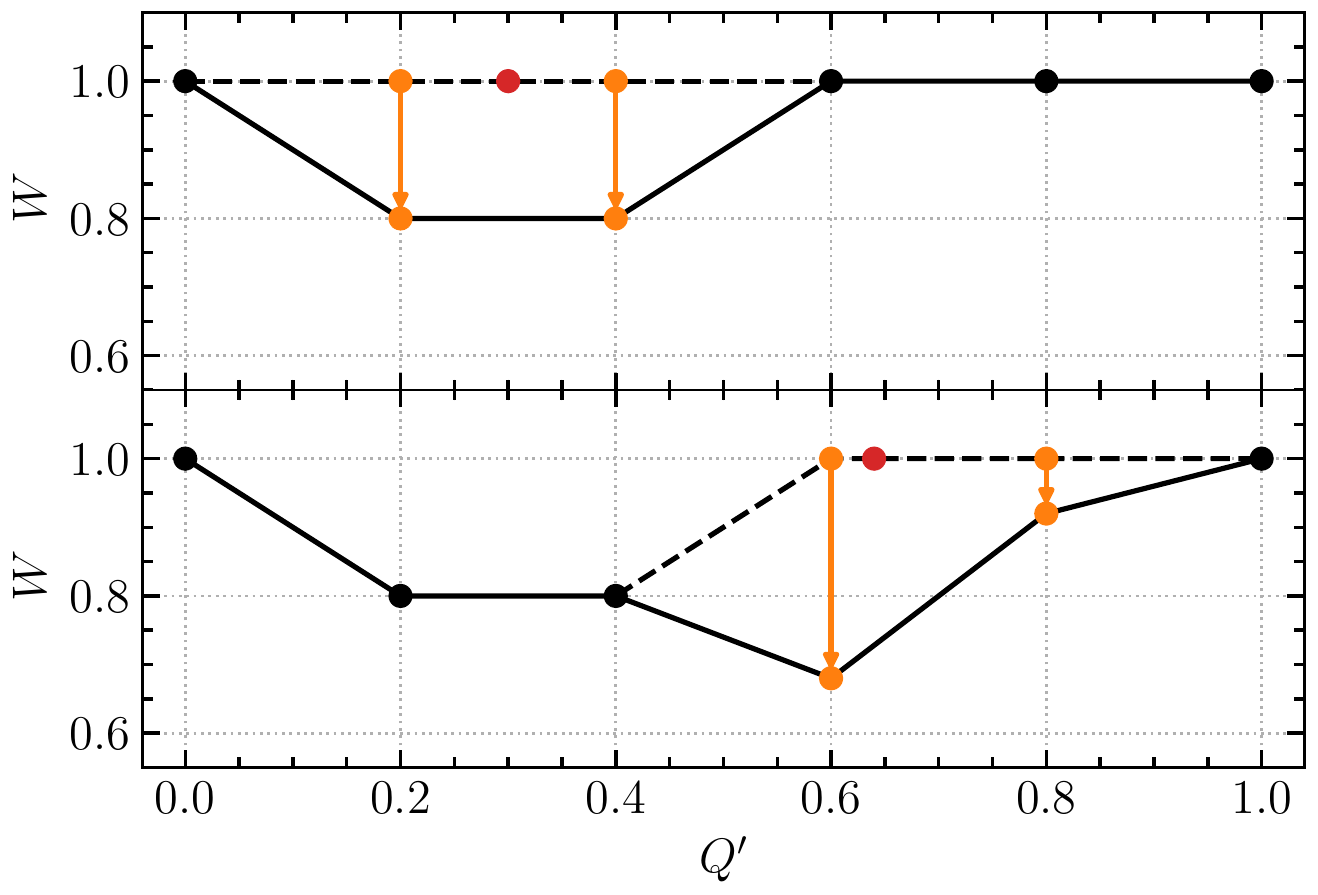}
     	\caption{Sketch of how the reweighting function is built for $N_\mathrm{int}=5$.
	The red points indicate the measurement of $Q'$ and the orange points the edges
	of the corresponding interval. The dashed line shows $W$ before the update, the solid line
	shows the updated $W$.}
     	\label{fig:Wbuild}
\end{figure}

In this first, preparatory Markov chain, we first need to define the topological sectors
that we want to include in the reweighting sample. As already discussed, at high
temperatures it is adequate to only regard the $Q=1$ sector and we
define the reweighting interval $\Omega_\mathrm{rew} = \left\{ Q' \in \left[ 0,1 \right] \right\}$.
The reweighting interval is further divided into $N_\mathrm{int}$ subintervals
$0 < Q'_1 < Q'_2 < \dots < Q'_\mathrm{int} = 1$ and $W$ is discretized on those points.
Between the interval borders, $W$ is kept piecewise linear, starting with a constant
function $W(Q') \equiv 1$. We then perform the reweighted HMC updates described above.
After each update, we assume that the current value of $Q'$ is oversampled, because the
algorithm managed to actually visit this $Q'$. Consequently, it should become less probable to
visit configurations with this $Q'$ again and $W$ is lowered close to the $Q'$ of the current configuration.
A sketch of this procedure is shown in Fig.~\ref{fig:Wbuild}. For more details, we refer to Ref.~\cite{Jahn:2018dke}.

\begin{figure}
	\centering
	\includegraphics[width=.6\textwidth]{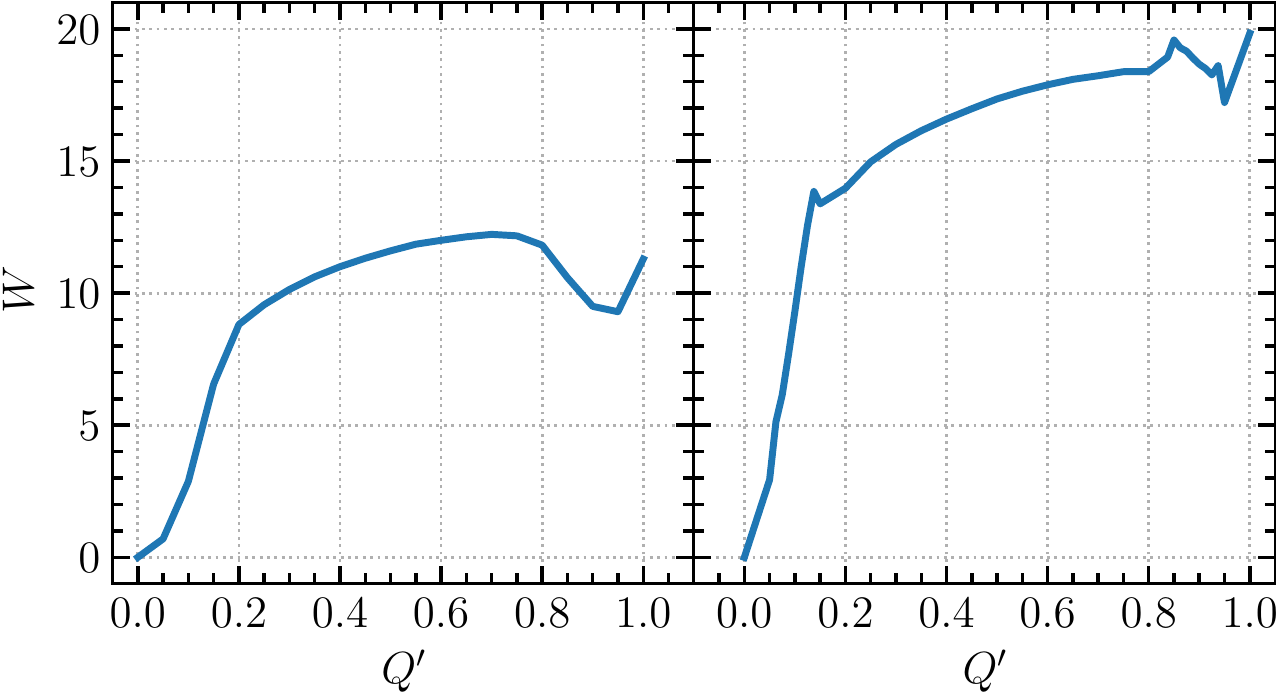}
     	\caption{Left: $W(Q')$ at $2.5~T_\mathrm c$. Right: $W(Q')$ at $4.1~T_\mathrm c$.}
     	\label{fig:Wfunc}
\end{figure}

In Fig.~\ref{fig:Wfunc}, the resulting reweighting function is shown for two different high temperatures above critical.
Both functions show local minima close to $Q'=0$ and $Q'=1$ indicating the presence of topological sectors.
In the intermediate $Q'$ range, the $W$ function remains large so as to enhance tunneling events between sectors.
Notice that  $\exp\left[-\left(W(Q'\simeq 1) - W(Q'\simeq 0)\right)\right]$ provides an approximate estimate for the
relative suppression of the $Q=1$ sector compared to the $Q=0$ sector and that this suppression is -- as expected --
much more severe for the higher temperature.

\newpage

\section{Results and Conclusions}
Ultimately we apply this strategy to SU(3) pure gauge theory at two
high temperatures, namely $2.5$ and $4.1~T_\mathrm c$. We use lattices
with three different lattice spacings using $N_\tau = 6,8,10$ and aspect ratios of about $2.5$.
For calculating the topological susceptibility via Eq.~\eqref{eq:chitop}, we need to determine
\begin{align}
\left\langle Q^2 \right\rangle \equiv \frac{\sum_i \e^{-W\left(Q'_i\right)} \uptheta\left(Q_i^2(t)-Q_\mathrm{thresh}^2\right)}{\sum_i \e^{-W\left(Q'_i\right)}},
\end{align}
where $Q_i(t)$ is the topological charge after a large amount $t$ of gradient flow such that all UV fluctuations
and dislocations are ``flowed away", and $Q_\mathrm{thresh}$ is a threshold to decide whether
the configuration is an instanton or not. We then compare different flow depth and different thresholds.

\begin{figure}[t]
	\centering
	\includegraphics[width=1\textwidth]{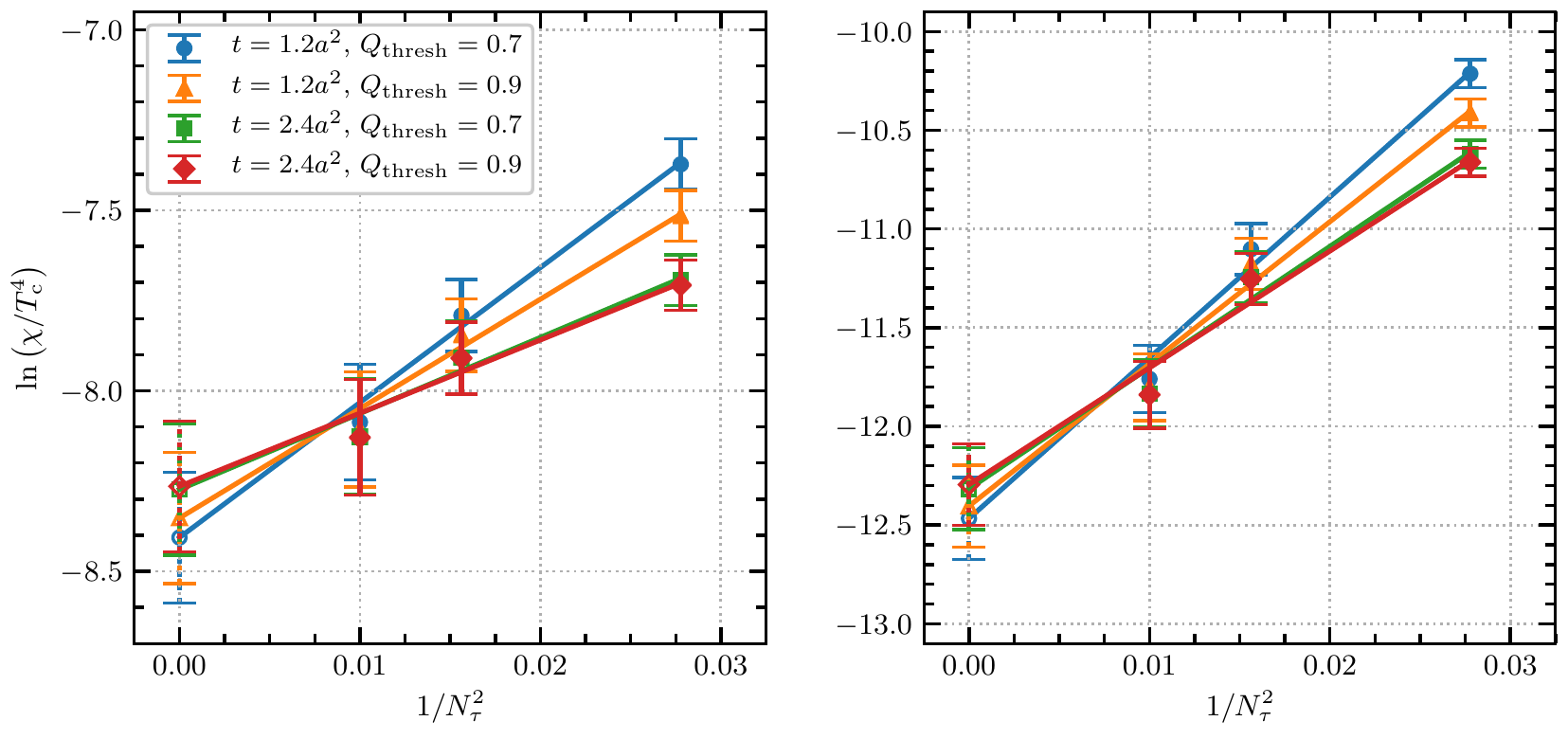}
     	\caption{Left: linear continuum extrapolation in the logarithm of the susceptibility for two different
	flow depth $t$ and two different thresholds $Q_\mathrm{thresh}$ at $T=2.5~T_\mathrm c$.
	Right: the same at $T=4.1~T_\mathrm c$.}
     	\label{fig:context}
\end{figure}

Fig.~\ref{fig:context} shows the continuum extrapolation at both temperatures for different
choices of flow depth and $Q_\mathrm{thresh}$. We see that at
coarse lattices the different topology definitions do not agree very well, while at finer lattices they do.
Also in the continuum limit, the different choices give the same result. The continuum
extrapolation is performed in the logarithm of the susceptibility because it is proportional to the
exponential of the caloron action which has $\mathcal O\left( a^2 \right)$ lattice corrections:
\begin{align}
\chi \propto \exp\left(-\left[1-\mathcal O\left(a^2T^2\right)\right] S\right).
\end{align}
From this, a linear extrapolation in the logarithm of the susceptibility is well justified \cite{Jahn:2018jvx}.
Our continuum extrapolated results are (using the $t=2.4a^2$ results with $Q_\mathrm{thresh}=0.7$)
\begin{align}
\begin{split}
\frac{\chi\left(T=2.5~T_\mathrm c\right)}{T_\mathrm c^4} &= 2.22 \times 10^{-4} \ \e^{\pm 0.18},
\\
\frac{\chi\left(T=4.1~T_\mathrm c\right)}{T_\mathrm c^4} &= 3.83 \times 10^{-6} \ \e^{\pm 0.21}
\end{split}
\end{align}
and agree well with the existing literature \cite{Berkowitz:2015aua,Borsanyi:2015cka}.

In conclusion, we developed a reweighting approach to measure the topological susceptibility
at high temperatures on the lattice. This method allows us to enhance tunneling between
topological sectors where standard lattice techniques fail, and lets us measure the topological
susceptibility directly also at high temperatures. We presented continuum extrapolated results
for SU(3) pure Yang-Mills theory and see no conceptual problems for applying the same
methods to QCD with fermions. Indeed, a similar technique was recently applied
to $N_\mathrm f = 2+1$ QCD \cite{Bonati:2018blm}.

\acknowledgments
The authors acknowledge support by the Deutsche Forschungsgemeinschaft
(DFG) through the grant CRC-TR 211  ``Strong-interaction matter under
extreme conditions". We also thank the GSI Helmholtzzentrum and the TU
Darmstadt and its Institut f\"ur Kernphysik for supporting this
research.

\bibliographystyle{JHEP}
\bibliography{refs}

\end{document}